\documentstyle[11pt,paspconf,epsf]{article}   

\def\G{$\Gamma_{\rm x}$ }
\def\ros{{\sl ROSAT }}
\def\asca{{\sl ASCA }}
\def\ein{{\sl Einstein }}

\def\0134{RX\,J0134-42}     
\def\7{1E\,0117.2-28}      
\def\approxlt{\mathrel{\hbox{\rlap{\lower.55ex \hbox {$\sim$}}
        \kern-.3em \raise.4ex \hbox{$<$}}}}
\def\approxgt{\mathrel{\hbox{\rlap{\lower.55ex \hbox {$\sim$}}
        \kern-.3em \raise.4ex \hbox{$>$}}}}

\begin{document}

\title{Warm absorbers in Narrow-Line Seyfert\,1 galaxies}
\author{Stefanie Komossa$^1$, Jochen Greiner$^2$}
\affil{$^1$ Max-Planck-Inst. f\"ur extraterrestrische Physik, 
   85740 Garching, Germany; ~ skomossa@xray.mpe.mpg.de \\
  $^2$ Astrophys. Institut Potsdam, 14482 Potsdam, Germany}

\begin{abstract}
Warm absorbers are an important new tool for investigating the conditions
within the central regions of active galaxies. 
They have been observed in $\sim$50\% of the well-studied Seyfert galaxies
and have also been detected in quite a number of Narrow-line Seyfert\,1 galaxies
(NLSy1). Here, we present a study of the X-ray properties of several 
NLSy1s with focus on their warm absorbers: 
(a) An analysis of all \ros PSPC observations of NGC\,4051
including new ones is performed, which reveals variability by a factor $\sim$30 in count rate 
and much less variability in the warm absorber parameters. 
(b) The possibility of a {\em dusty} warm absorber in IRAS 13349+2438
is explored on the basis of photoionization models for dusty warm gas 
and explicit \ros spectral fits.
(c) The X-ray spectrum of the NLSy1 1E\,0117.2-2837 is analyzed.
It can be successfully described by a very steep powerlaw of photon index
\G $\simeq -4$, or alternatively by a warm-absorbed flat powerlaw. UV-EUV
emission lines expected to arise from the warm material are predicted. 
(d) The strong spectral variability of RX J0134.3-4258 (from \G $\simeq -4.4$ 
in the \ros survey observation, to $\simeq -2.2$ in our subsequent pointing)
is examined in terms of warm absorption. 
\end{abstract}


\section{Introduction}

The presence of an ionized absorber was first discovered in \ein observations
of the quasar MR 2251-178 (Halpern 1984).
With the improved spectral resolution of \ros and {\sl ASCA}, many more were
found.  
So far, they revealed their existence mainly in the soft X-ray spectral region,
via absorption edges of highly ionized metal ions.
Their physical state,  location
and relation to other 
components of the active nucleus  
is still rather unclear.
E.g., an accretion disk wind and  
various BLR related models 
have been suggested.
Warm absorbers have been observed in $\sim$50\% of the well-studied
Seyfert galaxies and some quasars.
Characteristic X-ray absorption edges also show up in quite a number 
of NLSy1 galaxies (e.g., Leighly et al. 1996,
Wang et al. 1996, Guainazzi et al. 1996, 
Brandt et al. 1997, Komossa \& Fink 1997a, Hayashida 1997, Iwasawa et al. 1998)
and
we focus on this sub-group in the present contribution. We use the definition
of NLSy1 `loosely', i.e., we include the quasar IRAS\,13349+2438 which shares
some similarities with NLSy1s (Brandt et al. 1996, B96 hereafter).

\section{A {\itshape dusty} warm absorber: IRAS\,13349+2438 }

Recently, evidence has accumulated that some warm absorbers
contain significant amounts of dust (B96,
Komossa \& Fink 1996, Reynolds 1997, 
Komossa \& Fink 1997b-d, 
Reynolds et al. 1997, Leighly et al. 1997, Komossa \& Bade 1998,
Komossa 1998). 
The possibility  
of a {\em dusty} warm absorber in IRAS\,13349 was 
suggested by B96
to explain the lack of excess X-ray {\em cold} absorption despite strong optical
reddening 
(for follow-up \asca observations see Brinkmann et al. 1996,
Brandt et al. 1997).  
As emphasized by Komossa \& Fink (KoFi hereafter, e.g. 1997a,b) and 
Komossa \& Bade (KoBa, 1998) 
the influence of the presence of dust on the
X-ray absorption spectrum can be strong and becomes drastic for high 
column densities $N_{\rm w}$. 
Signatures of the presence of (Galactic-ISM-like) dust 
are, e.g., a strong carbon edge in the X-ray spectrum, 
and a stronger temperature gradient across the absorber
with more gas in a `cold' state. 

Here, we 
apply the model of a {\em dusty} warm absorber to the \ros X-ray
spectrum of IRAS 13349. Although repeatedly suggested, such a model
has not been fit previously (for some results see Komossa 1998).
Given the potentially strong modifications of the X-ray absorption spectrum
in the presence of dust, 
it is important to  scrutinize whether a dusty warm absorber is consistent with
the observed X-ray spectrum.
Since some strong
features of dusty warm absorbers appear outside the \asca sensitivity range,
\ros data are best suited for this purpose; we used the pointed PSPC observation
of Dec. 1992.
As ionizing continuum illuminating the warm absorber we adopted a mean Seyfert 
spectrum consisting of piecewise powerlaws 
(as in KoFi 1997b) with energy index $\alpha_{\rm EUV}=-1.4$. We use as definition
for the ionization parameter $U = Q/(4\pi{r}^{2}n_{\rm H}c)$. The photoionization calculations
were carried out with Ferland's (1993) code {\em Cloudy}.   

In a first step, we fit a dust-{\em free} warm absorber (as in B96,
but using the additional
information on the hard X-ray powerlaw available from the ASCA observation,
$\Gamma_{\rm x}^{\rm 2-10 keV} \simeq -2.2$).
This gives an excellent fit with log $N_{\rm w}$=22.7 ($\chi^2_{\rm red}$ = 0.8).
If this same model is re-calculated by fixing $N_{\rm w}$ and
the other best-fit parameters but {\em adding dust} to the warm absorber
the X-ray spectral shape is drastically altered and the data can not be
fit at all ($\chi^2_{\rm red}$ = 150). This still holds if we  
allow for
non-standard dust, i.e., selectively exclude either the graphite or silicate
species.

It has to be kept in mind, though, that the expected column derived
from optical extinction is less than the X-ray value of $N_{\rm w}$ determined
under the above assumptions. Therefore, in a next step, we
allowed all parameters (except \G) to be free and checked, whether a dusty
warm absorber could be fit.
This is not the case (e.g., if $N_{\rm w}$ is fixed to log $N_{\rm opt}$ = 21.2
we get $\chi^2_{\rm red}$ = 40).

The bad fit results can be partially traced back to the `flattening' effect of dust
(cf. Fig. 4 of KoBa 1998).
In fact, if we allow for a steeper intrinsic powerlaw spectrum, with \G $\simeq -2.9$
much steeper than the \asca value,
a dusty warm absorber with $N_{\rm w}=N_{\rm opt}$, the value derived 
from optical reddening, fits the \ros
spectrum well ($\chi^2_{\rm red}$ = 1.2, Tab. 1).
We also analyzed the \ros survey data and find the same
trends; i.e., the requirement of a steep intrinsic \ros spectrum.  
At present,
there are several possible explanations for the {\sl ROSAT}-\asca spectral differences:
(i) variability in a {\em two}-component warm absorber, (ii) variability in
the intrinsic powerlaw, or (iii) remaining {\sl ROSAT}-\asca
inter-calibration uncertainties.   

The presence of a {\em dusty} warm absorber in the NLSy1-like
quasar IRAS 13349 and the NLSy1 galaxy IRAS 17020+4544 (Leighly et al. 1997, KoBa 1998) 
further adds to the spectral complexity in NLSy1 galaxies.
Whereas early NLSy1 models tried to explain their very {\em steep}
observed X-ray spectra
by only one component (either a strong soft excess, or a warm absorber, or an intrinsically
steep spectrum), there is now evidence that
often all three components
are simultaneously present, and the additional presence of {\em dusty}
material partly compensates the `steepening effect' of the other three.

The detection of dust absorption features 
in the X-ray spectra of active galaxies with future X-ray missions will be an important
check and confirmation of the existence of {\em dusty} warm absorbers, and will
provide an
interesting approach to investigate dust properties in other galaxies.
But not all warm absorbers contain dust.

\begin{table} 
  \caption{X-ray  warm absorber (WA) fits to IRAS\,13349.}
 \begin{center}
  \begin{tabular}{llcllll}
  \tableline
      \noalign{\smallskip}
     & \multicolumn{4}{l}{warm absorber} & \multicolumn{2}{l}{single powerlaw} \\
         & \G~~~ & log $U$~~ & log $N_{\rm w}$~~ & $\chi^2_{\rm red}$~~~~~ &
                                              \G~~~ & $\chi^2_{\rm red}$ \\
      \noalign{\smallskip}
  \tableline
      \noalign{\smallskip}
    dusty WA  & --2.9 & --0.4 & 21.2$^{(1)}$ & 1.2 & --2.8 & 1.3 \\
    dust-free WA  & --2.2 &   ~0.7 & 22.7 & 0.8 & & \\
  \tableline
     \end{tabular}
  \label{tab1}

  \noindent{ \footnotesize $^{(1)}$ fixed to 
     the value $N_{\rm opt}$ determined from optical reddening }
\end{center}
\end{table}

\section {Dust-free warm absorber: NGC\,4051}

The NLSy1 galaxy NGC\,4051 
harbours a warm absorber
of high column density.
In this case, only a dust-free warm absorber fits the X-ray spectrum.
This can be traced back to the strong modifications of the
absorption structure in the presence of dust.{\footnote {We find 
hints in the X-ray spectrum of NGC\,4051 for a second, more lowly ionized
warm absorber that may be dusty, but better-than-available spectral resolution
is needed to check this.}} 

During a deep PSPC X-ray observation of Nov. 1993 the flux of NGC\,4051 turned out to be
strongly variable, the warm absorption features remained constant. This can be used 
to derive a constraint on the density and location of the ionized material, which yields 
$n \approxlt 3\,10^{7}$cm$^{-3}$ and $r \approxgt 3\,10^{16}$\,cm 
(see KoFi 1997a for details).

To investigate the long-term trend in the variability of NGC\,4051, in count rate
as well as in ionization parameter $U$ and column density $N_{\rm w}$ 
of the warm absorber, we analyzed previously unpublished
\ros PSPC observations of NGC\,4051 and also re-analyzed data earlier presented in 
Mc\,Hardy et al. (1995).
We find that in the long term all features are variable,
except for the cold absorption which is
always consistent with the Galactic value within the error bars.
$U$ and $N_{\rm w}$ change by about a factor of 2. The slope of the
powerlaw remains rather steep. 
We detect large-amplitude variability by a factor $\sim$30 in count rate
within the total observing interval.  
The long-term X-ray lightcurve is shown in Fig. 1, the best-fit warm-absorber parameters
are given in Tab. 2.

 \begin{figure}[ht]
\epsfxsize=12.8cm
\epsfbox{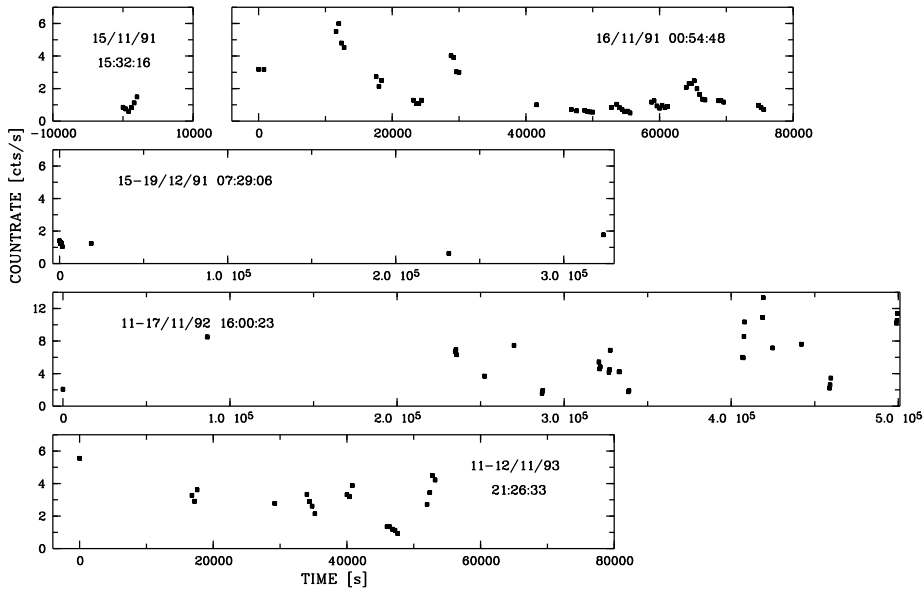}
 \caption[light]{Long-term X-ray lightcurve of NGC\,4051, based on all 
\ros PSPC observations. The source is variable by a factor $\sim$30 in count rate. 
The lightcurve of Nov.\,16,\,91 was earlier
shown in McHardy et al. (1995), the one of Nov.\,93 in
Komossa \& Fink (1997a). The time is measured in s from the beginnings of the individual
observations. }
\label{light}
\end{figure}

\section {Ultrasoft X-ray spectrum: 1E 0117.2-2837} 
This NLSy1 was discovered as an X-ray source by \ein and is
 at a redshift of $z$=0.347 (Stocke et al. 1991).
It is serendipituously located in one of the \ros PSPC pointings; the
steep X-ray spectrum was briefly noted by Schwartz et al. (1993). 
When described by a single powerlaw
continuum with Galactic cold column, the photon index is \G $\simeq -3.6$
(--4.3, if $N_{\rm H}$ is a free parameter).
The source's count rate, 0.44 cts/s, is constant throughout the observation. 

A successful description of the steep observed \ros spectrum 
is a warm-absorbed flat powerlaw
of canonical index.
We find a very large column density $N_{\rm w}$ in this case, and the contribution
of emission and reflection is no longer negligible; there is also some
contribution to Fe K$\alpha$. For the pure absorption model, the best-fit values
for ionization parameter and warm column density are $\log U \simeq 0.8$,
$\log N_{\rm w} \simeq 23.6$ ($N_{\rm H}$ is consistent with the  Galactic value),
with $\chi{^{2}}_{\rm red}$ = 0.74. Including the contribution of emission
and reflection for 50\% covering of the warm material gives
$\log N_{\rm w} \simeq 23.8$ ($\chi{^{2}}_{\rm red}$ = 0.65).

Several strong EUV emission lines are predicted to arise from the warm material.
Some of these are (H$\beta$ refers to the warm-absorber-intrinsic value):
FeXXI$\lambda$2304/H$\beta$ = 10, HeII$\lambda$1640/H$\beta$ = 16,
FeXXI$\lambda$1354/H$\beta$ = 37,
NeVIII$\lambda$774/H$\beta$ = 9, and, just outside the
IUE sensitivity range for the given $z$, FeXXII$\lambda$846/H$\beta$ = 113.
No absorption from CIV$\lambda$1549 and NV$\lambda$1240 is expected to show up in the UV spectrum. 
Both elements are more highly
ionized.

   \begin{table} 
  \caption{Log of \ros PSPC observations of NGC\,4051 and warm-absorber fit results.
                  $t_{\rm exp}$ = effective exposure time in s, $CR$ = mean count rate
                  in cts/s,
                  $L$ = {\em mean} (0.1--2.4 keV) luminosity corrected
                  for cold and warm absorption.}
      \begin{tabular}{ccccccc}
      \noalign{\smallskip}
      \hline
      \noalign{\smallskip}
       date & $t_{\rm exp}$ & $CR$ & log $U$ & log $N_{\rm w}$ &
              $\Gamma_{\rm x}$ & $L$ [10$^{41}$ erg/s] \\
      \noalign{\smallskip}
      \hline
      \noalign{\smallskip}
 Nov. 15, 1991 & ~2705 & 1.0 & 0.2 & 22.52 & -2.2 & 3.5\\
 Nov. 16, 1991 & 28727 & 1.6 & 0.2 & 22.45 & -2.2 & 5.4 \\
 Dec. 15 - 19, 1991 & ~4783 & 1.0 & 0.1 & 22.35 & -2.2 & 3.7 \\
 ~Nov. 11 - 17, 1992$^{*}$ & 20815 & 6.0 & 0.2 & 22.35 & -2.2 & 19.5~~\\
 Nov. 11 - 12, 1993 & 12261 & 2.9 & 0.4 & 22.67 & -2.3 & 9.5\\
      \noalign{\smallskip}
      \hline
      \noalign{\smallskip}
\end{tabular}

\noindent {\scriptsize
           $^{*}$ results of model fits uncertain due to off-axis location of source}
   \end{table}

\section {Strong spectral variability: RX\,J0134.3-4258}

This NLSy1 exhibited an ultrasoft spectrum in the \ros survey observation (Dec. 1990;
formally \G $\simeq -4.4$). Interestingly, the spectrum had changed to flat in
our subsequent pointed PSPC observation (Dec. 1992; \G $\simeq -2.2$).
This kind of spectral variability is naturally predicted within the framework of
warm absorbers, making the object a very good candidate for warm absorption.

We find that a warm-absorbed, intrinsically flat powerlaw can indeed describe the
survey observation.
Due to the low number of available photons, a range of possible combinations
of $U$ and $N_{\rm w}$ explains the data
with comparable success.
A large
column density $N_{\rm w}$ (of the order 10$^{23}$ cm$^{-2}$) is needed to
account for the ultrasoft observed spectrum.
The most suggestive scenario within the framework of warm absorbers
is a change in the {\em ionization state} of ionized material along the line of sight,
caused by {\em varying irradiation} by a central ionizing source.
In the simplest case, lower intrinsic luminosity would be expected, in order to cause the deeper
observed absorption in 1990, whereas the source is somewhat brighter in the survey observation
(see Komossa \& Fink 1997d for details).  
Some variability seems to be usual, though: the count rate changed by
about a factor of 2 during the pointed observation (Fig. 2). If one wishes to keep this
scenario, one has to assume that the ionization state of the absorber still reflects
a preceding (unobserved) low-state in intrinsic flux.
Alternatively, gas heated by the central continuum source
may have {\em crossed the line of sight},
producing the steep survey spectrum, and has (nearly) disappeared in the 1992 observation.
Finally, we note that an alternative explanation of the spectral variability in terms of a variable
soft excess was suggested by Grupe (1996); see also Mannheim et al. (1996).  

 \begin{figure} 
\epsfxsize=6.5cm
\epsfbox{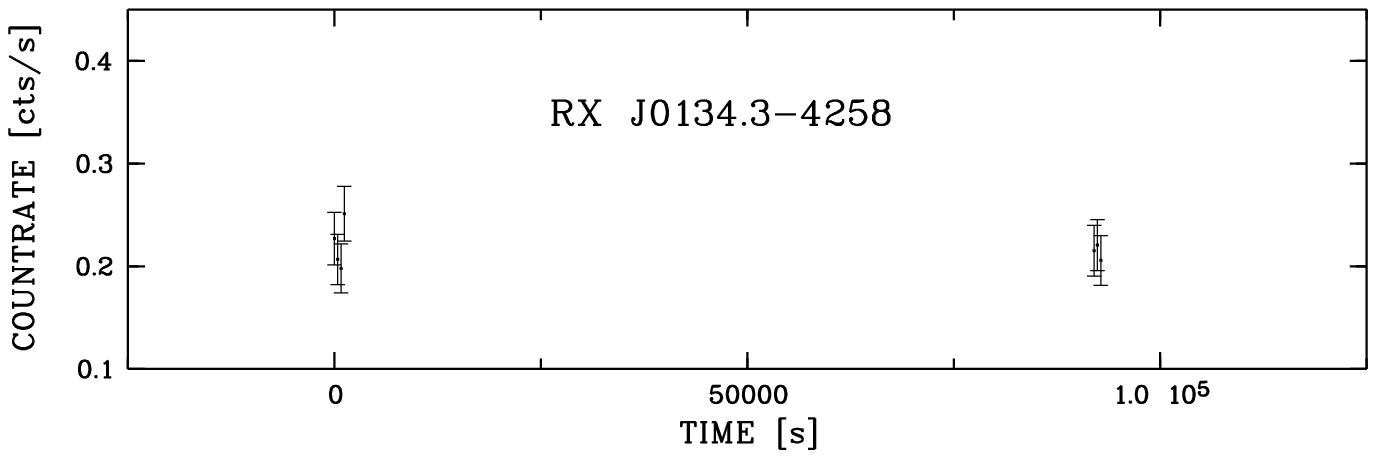}
    \vspace*{-2.3cm}\hspace*{7.0cm}
\epsfxsize=6.5cm
\epsfbox{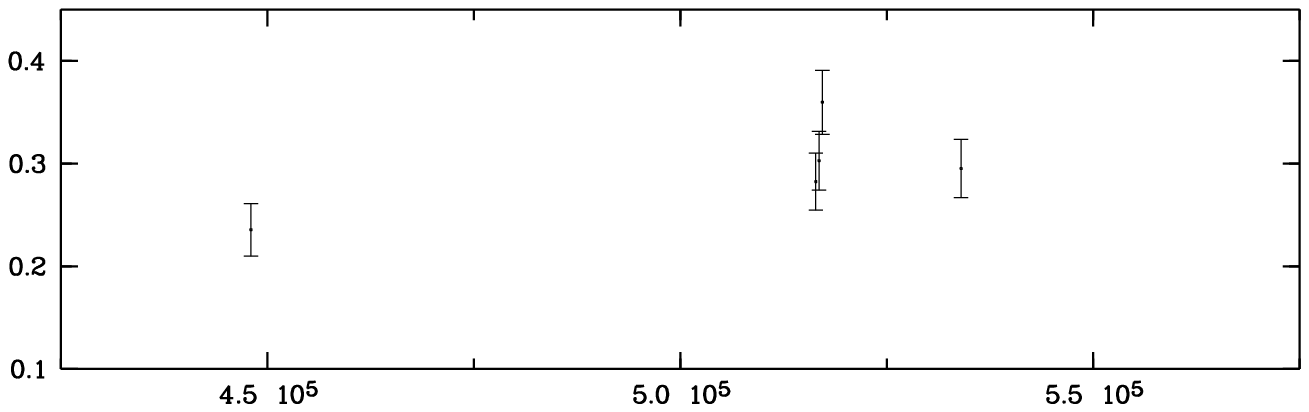}
 \caption[light]{X-ray lightcurve of \0134 during the pointed \ros PSPC
observation. The data were binned to time intervals of 400s.} 
\label{light}
\end{figure}

\acknowledgments
St.K. and J.G. acknowledge support from the Verbundforschung under grants 
No. 50\,OR\,93065 and 50 QQ 9602 3, respectively.
We thank Gary Ferland for providing {\em Cloudy}.

\end{document}